\documentclass[twocolumn,showpacs,superscriptaddress,floatfix,prl]{revtex4-1}
\usepackage{graphicx,amsfonts,amssymb,amsmath,hyperref,hypcap,enumerate}

\usepackage{amsthm}
\usepackage{multirow}
\usepackage{graphicx}
\usepackage{dcolumn}   
\usepackage{bm}        
\usepackage{amssymb}   
\usepackage{amsmath}
\usepackage{epstopdf}
\usepackage{color}

\DeclareMathAlphabet{\mathitb}{OT1}{cmr}{bx}{sl}

\begin{document}
	\title{Critical Behaviors of Anderson Transitions in Three Dimensional Orthogonal Classes with Particle-hole Symmetries}

\author{Xunlong Luo}
\affiliation{International Center for Quantum Materials, Peking University, Beijing 100871, China}
\affiliation{Collaborative Innovation Center of Quantum Matter, Beijing 100871, China}

\author{Baolong Xu}
\affiliation{International Center for Quantum Materials, Peking University, Beijing 100871, China}
\affiliation{Collaborative Innovation Center of Quantum Matter, Beijing 100871, China}

\author{Tomi Ohtsuki}
\affiliation{Department of Physics, Sophia University, Chiyoda-ku, Tokyo 102-8554, Japan}

\author{Ryuichi Shindou}
\email{rshindou@pku.edu.cn}
\affiliation{International Center for Quantum Materials, Peking University, Beijing 100871, China}
\affiliation{Collaborative Innovation Center of Quantum Matter, Beijing 100871, China}

\date{\today}
	\begin{abstract}
From transfer-matrix calculation of localization lengths and their finite-size scaling 
analyses, we evaluate critical exponents of the Anderson  
metal-insulator transition in three dimensional (3D) orthogonal class with
particle-hole symmetry, class CI, as $\nu=1.16\pm 0.02$.
We further study disorder-driven quantum phase transitions in the 3D nodal 
line Dirac semimetal model, which belongs to class BDI, and estimate critical exponent as 
$\nu=0.80\pm 0.02$. From a comparison of the critical exponents, 
we conclude that a disorder-driven re-entrant insulator-metal transition 
from the topological insulator phase in the class BDI to the diffusive 
metal phase belongs to the same universality class as the 
Anderson transition in the 3D class BDI. 
We also argue that an infinitesimally small disorder drives the nodal line Dirac semimetal
in the clean limit to the diffusive metal. 
	\end{abstract}
\maketitle
\textit{Introduction} --- Identifying a new critical behavior in quantum
phase transition is one of the fundamental subject in physics. In theory, 
critical exponent and scaling function represent universal aspects of a saddle-point fixed 
point of an underlying renormalization group (RG) equation~\cite{Cardy96}. 
In Anderson metal-insulator transition~\cite{Anderson58}, these quantities are key 
ingredients of universal scaling properties of electric and thermal transports around the quantum 
phase transition, and they are determined only by the basic symmetries and spatial dimension 
of Hamiltonian~\cite{Wegner76,Abrahams79,Efetov80,hikami81,Pichard81,MacKinnon81,MacKinnon83}. 
Recent material discoveries of Weyl~\cite{Xu15science,Lv15,Yan17,Armitage18} and 
nodal line Dirac semimetals~\cite{Lou16,Takane16,Bian16,Schoop16,Hosen17,Wang17,Liu18} stimulate intensive 
studies on non-Anderson-type disorder-driven quantum phase transitions~\cite{Syzranov15,Syzranov15PRL,Syzranov18}. 
Their universal critical properties are characterized by 
unconventional critical exponents~\cite{Kobayashi14,Sbierski14,Sbierski15,Pixley15,Pixley16,Liu16,Luo_quantumMulti} 
and scaling forms~\cite{Luo_Unconventional}, indicating new universality classes 
of the disorder-driven quantum phase transitions.  

In this paper, we evaluate numerically critical exponents of the Anderson 
transitions in three dimensional (3D) systems with particle-hole symmetries, 3D symmetry  
class CI and symmetry class BDI~\cite{Gade91,Gade93,Altland97,Schnyder08}, respectively. 
We study disorder-driven metal-insulator transitions in a nodal line
Dirac semimetal model in the class BDI. 
From a comparison of the critical exponents, we conclude 
that a disorder-driven re-entrant transition between topological band insulator 
phase in the class BDI and diffusive metal phase belongs to the 
same universality class~\cite{Obuse07,Fu12} 
as the Anderson transition in the 3D class BDI~\cite{Garcia06}. 
Contrary to a previous study~\cite{Gonalves19} where nodal line
Dirac semimetal is stable against disorder, we argue that an infinitesimally 
small disorder drives the nodal line Dirac semimetal to the diffusive metal.

\begin{table*}[t]
	\begin{tabular}{c|c|c|c|c|c|c|c|c}
		\hline 
		phase transition label& $m_1$ & $n_1$ & $m_2$ & $n_2$ & GOF & $W_c$ & $\nu$ & $y$ \\ 
		\hline 
3D class CI & 2 & 3 & 0 & 1 & 0.12 & 
10.957 \!\ [10.953 , 10.961] & 1.160 \!\ [1.144 , 1.174] & 1.20 \!\ [0.93, 1.74] \\ 
3D class CI & 3 & 3 & 0 & 1 & 0.11 & 
10.957 \!\ [10.953 , 10.961] & 1.160 \!\ [1.142 , 1.176] & 1.21 \!\ [0.96, 3.09] \\ \hline
	phase transition 1 \!\ (3D class BDI) \!\ &3&3&0&1&0.12
&3.135 \!\ [3.132 , 3.138]& 0.832 \!\ [0.723 , 0.906]& 2.95 \!\ [1.80 , 4.49]\\
	phase transition 2 \!\ (3D class BDI) \!\ &3&3&0&1&0.28
&11.96 \!\ [11.92 , 12.02]& 0.798 \!\ [0.753 , 0.832]& 1.45 \!\ [1.25 , 1.66]\\
    phase transition 3 \!\ (3D class BDI) \!\ &2&3&0&1&0.18
&4.76 \!\ [4.75 , 4.77]& 0.824 \!\ [0.803 , 0.846]& 3.35 \!\ [2.28 , 4.21]\\
	phase transition 3 \!\ (3D class BDI) \!\ &3&3&0&1&0.18&
4.76 \!\ [4.75 , 4.77]& 0.825 \!\ [0.800 , 0.846]& 3.33 \!\ [2.40 , 4.21]\\
		\hline 
	\end{tabular} 
	\caption{Polynomial fitting results for the metal-insulator transitions measured in the 3D class CI model 
         and at three different sets of parameters in the 3D class BDI model. In the class BDI model, the phase 
         transition 1 is from topological band insulator to diffusive metal, the phase transition 2 is from 
         diffusive metal to Anderson insulator, and the phase transition 3 is from trivial band insulator to 
         diffusive metal (see in FIG.~\ref{fig:1}). 
         The goodness of fit (GOF), critical disorder $W_{c}$, critical exponent $\nu$, 
         the scaling dimension of the least irrelevant scaling variable $-y$ are shown for different 
         orders of the Taylor expansion of the universal scaling function $(m_1,n_1,m_2,n_2)$. 
         The square bracket is the 95\% confidence interval.}
	\label{table:1}
\end{table*}

\textit{3D class CI} --- Let us begin with 3D tight-binding model belonging to class CI. The 
following two-orbital cubic-lattice model is considered in this paper; 
\begin{align}
{\cal H} \equiv& \sum_{{\bm i},{\bm j}} \sum_{d,d^{\prime}}
| {\bm i},d \rangle
[\mathbb{H}]_{({\bm i},d|{\bm j},d^{\prime})} \langle{\bm j},d^{\prime}|  \nonumber \\
 = &\sum_{{\bm i}} \Big\{ \big(\varepsilon_{{\bm i}} + \Delta \big) \big(
|{\bm i},a\rangle 
\langle {\bm i},a| - |{\bm i},b\rangle   \langle {\bm i},b| \big) \nonumber \\
&   + t_{\parallel}  \big(|{\bm i},a\rangle 
\langle {\bm i},b| + |{\bm i},b\rangle   \langle {\bm i},a| \big) \nonumber \\   
& 
 + t_{\perp} \sum_{\mu=x,y} \sum_{d=a,b} \big(|{\bm i}+{\bm e}_{\mu},d\rangle 
+ |{\bm i}-{\bm e}_{\mu},d\rangle  \big) \langle {\bm i},d| \nonumber \\  
&   + t^{\prime}_{\parallel} \big(|{\bm i}+{\bm e}_z,a\rangle  \langle {\bm i},a| 
- |{\bm i}+{\bm e}_z,b\rangle  \langle {\bm i},b| + {\rm h.c.}  \big) \Big\}. \label{3dci}
\end{align} 
Here $d,d^{\prime}=a,b$ denotes the orbital index, ${\bm i}\equiv (i_x,i_y,i_z)$ with 
${\bm e}_x=(1,0,0)$, ${\bm e}_y=(0,1,0)$ and ${\bm e}_z=(0,0,1)$ is the site index 
on the 3D cubic lattice. $\varepsilon_{\bm i}$ represents a random potential, which is uniformly distributed 
in a range of $[-W/2,W/2]$. The random potentials at two different 
cubic lattice sites have no correlation; $\overline{ \varepsilon_{\bm i} \varepsilon_{\bm j} } 
= \delta_{{\bm i},{\bm j}} W^2/12$. The model with the random potential 
has a particle-hole symmetry (${\mathbb{P}} {\mathbb{H}} {\mathbb{P}} = - {\mathbb{H}}$) as well as the 
time-reversal symmetry (${\mathbb{H}}^* = {\mathbb{H}}$) with 
$[{\mathbb{P}}]_{({\bm i},d|{\bm j},d^{\prime})} \equiv (-1)^{i_x+i_y} \delta_{{\bm i},{\bm j}} 
[\sigma_y]_{d,d^{\prime}}$. Since ${\mathbb{P}}^{\rm T}= - {\mathbb{P}}$, 
the single-particle Hamiltonian has a set of 
doubly degenerate real-valued eigenstates at the zero eigenenergy, 
which results in the degeneracy of the Lyapunov exponents at $E=0$; 
the degeneracy is protected by the particle-hole symmetry. According to the symmetry classification of 
the random matrix theory~\cite{Altland97}, the zero-energy eigenstates of ${\mathbb{H}}$ 
belong to the class CI. In the following, we set $\Delta=t_{\parallel}=t^{\prime}_{\parallel}=t_{\perp}=1$ 
and focus on a delocalization-localization transition of the zero-energy eigenstates. In the clean limit 
($W=0$), the Hamiltonian has two disconnected Fermi surfaces at $E=0$~\cite{supplemental}. 
In the presence of the disorder, a localization length of the zero-energy eigenstates along the $z$-direction 
($\lambda_z$) is calculated in terms of the transfer matrix 
method~\cite{Pichard81,MacKinnon81,MacKinnon83,Slevin14}. The 
periodic boundary condition is imposed along $x$ and $y$ directions with a linear dimension 
within the $xy$ plane ($L$). On increasing the disorder strength $W$, the eigenstates 
at $E=0$ undergo the Anderson transition. The quantum phase transition is 
detected by a scale-invariant behavior of a normalized 
localization length $\Lambda_z \equiv \lambda_z/L$~\cite{supplemental}. 
The density of states (DOS) of ${\mathbb{H}}$ with 
finite disorder strength $W$ is calculated in terms of kernel polynomial expansion (KPE) 
method~\cite{Weisse06}. Due to the particle-hole symmetry, 
the calculated DOS is symmetric about $E=0$, while the DOS at $E=0$ remains 
finite at the quantum phase transition point~\cite{supplemental}.

The critical exponent of the Anderson transition in the 3D class CI model 
is determined by polynomial fitting method~\cite{Slevin14}. Under an 
assumption of spatially isotropic scaling property of a saddle-point fixed 
point, the normalized localization length $\Lambda_z$ should be given by a 
scaling function $\Lambda_z = F(\phi_1,\phi_2)$ where 
$\phi_1 \equiv u_1(w) L^{1/\nu}$ and $\phi_2 \equiv u_2(w) L^{-y}$ stand for 
a relevant and irrelevant scaling variable at the saddle-point fixed point; $1/\nu \!\ (>0)$ and 
$-y \!\ (<0)$ are the scaling dimensions of the relevant and irrelevant scaling variables around the postulated 
saddle-point fixed point. $w$ is a normalized distance from the critical point; $w \equiv (W-W_c)/W_c$. 
When $W$ is sufficiently close to the critical disorder strength $W_c$, $u_{1}(w)$ and $u_{2}(w)$ can be Taylor 
expanded in small $w$. By definition, the expansions take forms of 
$u_{i}(w) \equiv \sum^{m_i}_{j=0} b_{i,j} w^{j}$with $i=1,2$, 
$b_{1,0}=0$ and $b_{2,0} \ne 0$. For smaller $w$ and larger $L$, the universal function 
can be further expanded in small $\phi_1$ and $\phi_2$ as 
$F = \sum^{n_1}_{j_1=0} \sum^{n_2}_{j_2=0} a_{j_1,j_2} \phi^{j_1}_1 \phi^{j_2}_2$.
For a given set of $(n_1,n_2,m_1,m_2)$, $\chi^2 \equiv \sum^{N_D}_{k=1} (F_{k}-\Lambda_{z,k})^2/\sigma^2_{k}$ 
is minimized in terms of $W_c$, $\nu$, $-y$, $a_{i,j}$ and $b_{i,j}$ (without loss of generality, we set 
 $a_{1,0}=a_{0,1}=1$). $N_D$ here is a number of data points used for the fitting, and
each data point $k$ is specified by $L$ and $W$. $\Lambda_{z,k}$ and $\sigma_k$ are a mean value and 
error bar of $\Lambda_z$ from the transfer matrix calculation for $k=(L,W)$, 
respectively, while $F_{k}$ is a fitting value from the polynomial expansion of the universal function $F$ 
for the same $L$ and $W$. Fittings are carried out for several different $(n_1,n_2,m_1,m_2)$ with 
$n_1\le 3$, $n_2=1$, $m_1\le 3$ and $m_2=0$. Table~\ref{table:1} shows the fitting results with 
goodness of fit (GOF) greater than 0.1. The same fittings are also carried out for 1000 sets of $N_D$ 
number of synthetic data that are generated from the mean value and the error bar 
at each data point. The fittings for the synthetic data give 95 $\%$ confidence intervals 
in Table~\ref{table:1}. From the polynomial fitting analyses, 
the critical exponent of the Anderson transition in the 
3D class CI is evaluated as $\nu=1.16 \pm 0.02$. The critical 
exponent thus evaluated is clearly distinct from any of the conventional critical exponents in the 
Wigner-Dyson universality classes in 3D~\cite{Slevin14,Slevin97,Asada05}.

\textit{3D class BDI} --- Let us next introduce a 3D tight-binding model in the class BDI, 
whose clean limit encompasses nodal line Dirac semimetal as well as topological band insulator phases. 
The following two-orbital cubic-lattice model is considered;
\begin{align}
{\cal H} \equiv &\sum_{{\bm i},{\bm j}} \sum_{d,d^{\prime}}
| {\bm i},d \rangle [{\mathbb{H}}]_{({\bm i},d|{\bm j},d^\prime)}
\langle{\bm j},d^{\prime}|  \nonumber \\
= & \sum_{{\bm i}} \bigg[ \big(\varepsilon_{{\bm i}} + \Delta \big) \big(|{\bm i},a\rangle 
\langle {\bm i},a| - |{\bm i},b\rangle  \langle{\bm i},b| \big) \nonumber \\
&  
+ t_{\parallel} \Big\{
\big(|{\bm i}+{\bm e}_z,a\rangle - |{\bm i}-{\bm e}_z,a\rangle\big) 
 \langle{\bm i},b| + {\rm h.c.} \big\} \nonumber \\   
& 
 + t_{\perp} \sum_{\mu=x,y}  \big(|{\bm i}+{\bm e}_{\mu},a\rangle 
\langle {\bm i},a| - |{\bm i}+{\bm e}_{\mu},b\rangle \langle{\bm i},b| + {\rm h.c.} \big) \nonumber \\  
&  + t^{\prime}_{\parallel} \big(|{\bm i}+{\bm e}_z,a\rangle \langle{\bm i},a| 
- |{\bm i}+{\bm e}_z,b\rangle \langle{\bm i},b| + {\rm h.c.}  \big) \bigg], \label{3dbdi}
\end{align} 
where the same notation as in Eq.~(\ref{3dci}) is used. 
The model with non-zero disorder has a time-reversal symmetry  
(${\mathbb{H}}^{*}={\mathbb{H}}$) and a particle-hole symmetry 
(${\mathbb{P}}^{\prime} {\mathbb{H}} {\mathbb{P}}^{\prime} = - {\mathbb{H}}$) 
with $[{\mathbb{P}}^{\prime}]_{({\bm i},d|{\bm j},d^{\prime})}=\delta_{{\bm i},{\bm j}} [\sigma_x]_{d,d^{\prime}}$. 
Since ${{\mathbb{P}}^{\prime}}^{\rm T} = {\mathbb{P}}^{\prime}$, 
the zero-energy eigenstates of ${\mathbb{H}}$ as well as the Lyapunov exponents 
have no symmetry-protected degeneracy. 
According to the symmetry classification~\cite{Altland97}, the zero-energy eigenstates 
belong to the class BDI. We emphasize that compared to a bipartite-lattice model  
with hopping disorders~\cite{Garcia06}, the potential disorder that preserves 
the particle-hole symmetry enables a stable transfer matrix calculation of the localization 
length and evaluation of the critical exponent in the 3D class BDI.  In the following, 
we set $t^{\prime}_{\parallel}=-1$, $t_{\parallel}=-1/4$  and either $t_{\perp}=3/10$  
varying $\Delta$ and $W$ or $\Delta=0$ varying $t_{\perp}$ and $W$, and focus 
on the quantum phase transitions of the zero-energy eigenstates. 

\begin{figure}
	\centering
	\includegraphics[width=1\linewidth]{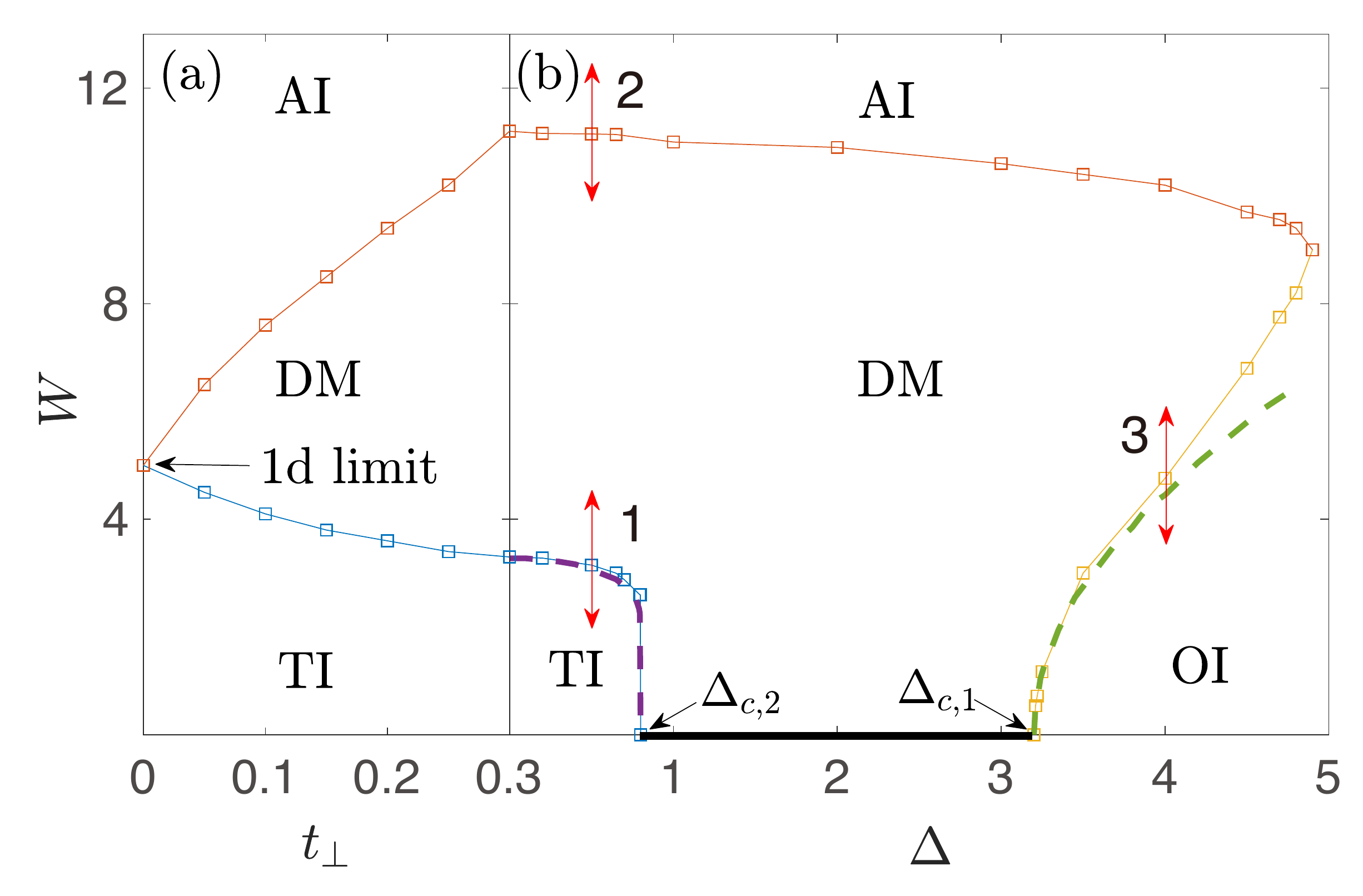}
	\caption{Phase diagram of the 3D class BDI model determined by the transfer matrix calculation 
		of the localization length along $x$ direction ($t_{\perp} \ne 0$) and along $z$ direction 
        ($t_{\perp}=0$) and by the self-consistent Born calculation (dotted line). 
        (a) $W$-$t_{\perp}$ diagram ($\Delta=0$) (b) $W$-$\Delta$ diagram ($t_{\perp}=3/10$). 
        The boundary by the square boxes is the metal-insulator 
		transition line of the zero-energy states, while the dotted line is a critical disorder strength above 
		which the zero-energy density of state becomes finite. 
		The bold black line represents nodal line Dirac semimetal phase. 
		The red double-headed arrows stand for the phase transitions we focus on in this paper.
		AI, DM, TI and OI stand for Anderson 
		insulator, diffusive metal, topological band insulator and trivial band insulator phases, 
        respectively (see the text).}
	\label{fig:1}
\end{figure}

In the clean limit ($W=0$), an energy-momentum dispersion of ${\mathbb{H}}$ is 
given by $E_{\pm}({\bm k}) = \pm \{[\Delta+ 2t_{\perp}(c_x+c_y)+
2t^{\prime}_{\parallel} c_z]^2 + 4t^2_{\parallel} s^2_z\}^{\frac{1}{2}}$ 
where $c_{\mu} \equiv \cos k_{\mu}$, $s_{\mu} \equiv \sin k_{\mu}$ ($\mu=x,y,z$), 
${\bm k}\equiv (k_x,k_y,k_z)$, ${\bm H}({\bm k})|u_{\pm}({\bm k})\rangle = E_{\pm}({\bm k})|u_{\pm}({\bm k})\rangle$, and ${\bm H}({\bm k})$ is a Fourier-transform of 
${\mathbb{H}}$ without the random potential.  The two energy bands undergo a 
sequence of phase transitions as a function of $\Delta$. When $\Delta>\Delta_{c,1} \equiv 
2| t^{\prime}_{\parallel}| +4|t_{\perp}|=16/5$, there is a finite band gap between the two energy bands 
(trivial band insulator phase). When 
$\Delta_{c,1}>\Delta>\Delta_{c,2} \equiv 2|t^{\prime}_{\parallel}|-4|t_{\perp}|=4/5$, 
the two bands form a close loop of band touchings at $E=0$ in the momentum space, that lies on a plane of 
$k_z=0$ (nodal line Dirac semimetal phase). On decreasing $\Delta$, 
the loop grows up from $(k_x,k_y)=(\pi,\pi)$ at $\Delta=\Delta_{c,1}$
and shrink into $(0,0)$ at $\Delta=\Delta_{c,2}$. The two energy bands form a 
pair of two linearly dispersive Dirac cones within a cross-sectional plane that cut the closed 
loop into two open lines; the loop is called as Dirac nodal line. Due to the Dirac cone, 
the Zak phase~\cite{zak89} along 
the $k_z$ axis, $i \int^{\pi}_{-\pi} \langle u_{\pm} ({\bm k})|\partial_{k_z} |u_{\pm}({\bm k})\rangle \!\ d
k_z$, is $\pi$ and $0$ whenever $(k_x,k_y)$ is inside and outside the closed loop; $(k_x,k_y)=(\pi,\pi)$ / 
$(0,0)$ is inside/outside the loop. The $\pi$ Zak phase leads to Su-Schrieffer-Heeger (SSH) 
zero-energy states~\cite{Su79} localized at the spatial boundary along $z$. 
The SSH states form a `drum-head' shape zero-energy 
flat surface state in the surface Brillouin zone, where a boundary of the `drum head' is 
given by a projection of the  Dirac nodal line onto the surface BZ. 
When $\Delta<\Delta_{c,2}$, the two bands 
open a gap again, while the Zak phase is $\pi$ for 
any $(k_x,k_y) \in [-\pi,\pi] \times [-\pi,\pi]$; the spatial boundary 
along $z$ has the SSH zero modes~\cite{Su79} for any surface crystal 
momentum in the surface BZ (topological band insulator phase). 
The topological insulator phase is equivalent to the one dimenisonal (1D) 
topological insulator in the class BDI. Namely, by turning off $t_{\perp}$ in 
Eq.~(\ref{3dbdi}), one can adiabatically connect the topological band 
insulator phase to decoupled 1D models with a finite band gap at $E=0$.

\textit{Re-entrant insulator-metal transition} --- The 1D topological insulator in the class BDI is 
characterized by an integer-valued topological number ${\mathbb{Z}}$~\cite{Schnyder08}. 
In the clean limit, the integer corresponds to a winding number~\cite{Wen89} between the 1D 
Brillouin zone for $k_z$ and a loop formed by two Pauli matrices in ${\bm H}({\bm k})$. 
The winding number of the topological insulator phase in $\Delta<\Delta_{c,2}$ is 
$+1$~\cite{supplemental}. When the random pontential is 
weakly introduced with the BDI symmetry preserved, the topological integer 
remains unchanged, unless the zero-energy bulk states undergo a 
localization-delocalization transition. Meanwhile, the zero-energy 
states in strongly disordered regime should be in a localized phase with the zero topological 
integer (Anderson insulator phase). This indicates that between the topological band  
insulator phase in weakly disordered regime and the Anderson insulator phase in strongly 
disordered regime, there must be two-step disorder-driven quantum phase transitions: one 
transition from the topological band insulator  
to metal phases and the other from metal to the Anderson insulator phases. In the 
1D limit, a metal phase cannot exist in the presence of finite disorder. Thus, the two 
phase transition points must collapse into a point in the limit of 
$t_{\perp}\rightarrow 0$. The transfer matrix calculation of the localization 
lengths~\cite{supplemental} confirms this global structure of the 
phase diagram (Fig.~\ref{fig:1}).  

\textit{Effect of disorders in nodal line Dirac semimetal} --- The bulk DOS in 
the  nodal line Dirac semimetal (NLDSM) vanishes linearly in $E$ at the node ($E=0$) in 
the clean limit. When the random potential $\varepsilon_{\bm i}$ with a finite disorder strength is 
introduced, the bulk zero-energy states acquire a finite mean-free (life) time, making the DOS at 
$E=0$  finite. We call this metal phase with finite zero-energy DOS as diffusive metal (DM) 
phase and distinguish DM phase from NLDSM phase with the vanishing zero-energy DOS.

The short-ranged random potential is a marginally relevant scaling variable around the clean-limit 
fixed point (NLDSM fixed point), and an infinitesimally small disorder always transforms the NLDSM 
phase into DM phase. To see this, note first that the degeneracy between the two energy bands 
is not lifted in a tangential direction along the closed loop. 
As a result, a tree-level scaling dimension of the momentum along the tangential direction is 
zero. This makes the tree-level scaling dimension of the short-ranged disorder strength to be 
zero. Being given by an attractive interaction in an effective action, the quenched 
disorder strength is always reinforced by a one-loop RG correction around the clean-limit fixed point.

\begin{figure}[t]
	\centering
	\includegraphics[width=1\linewidth]{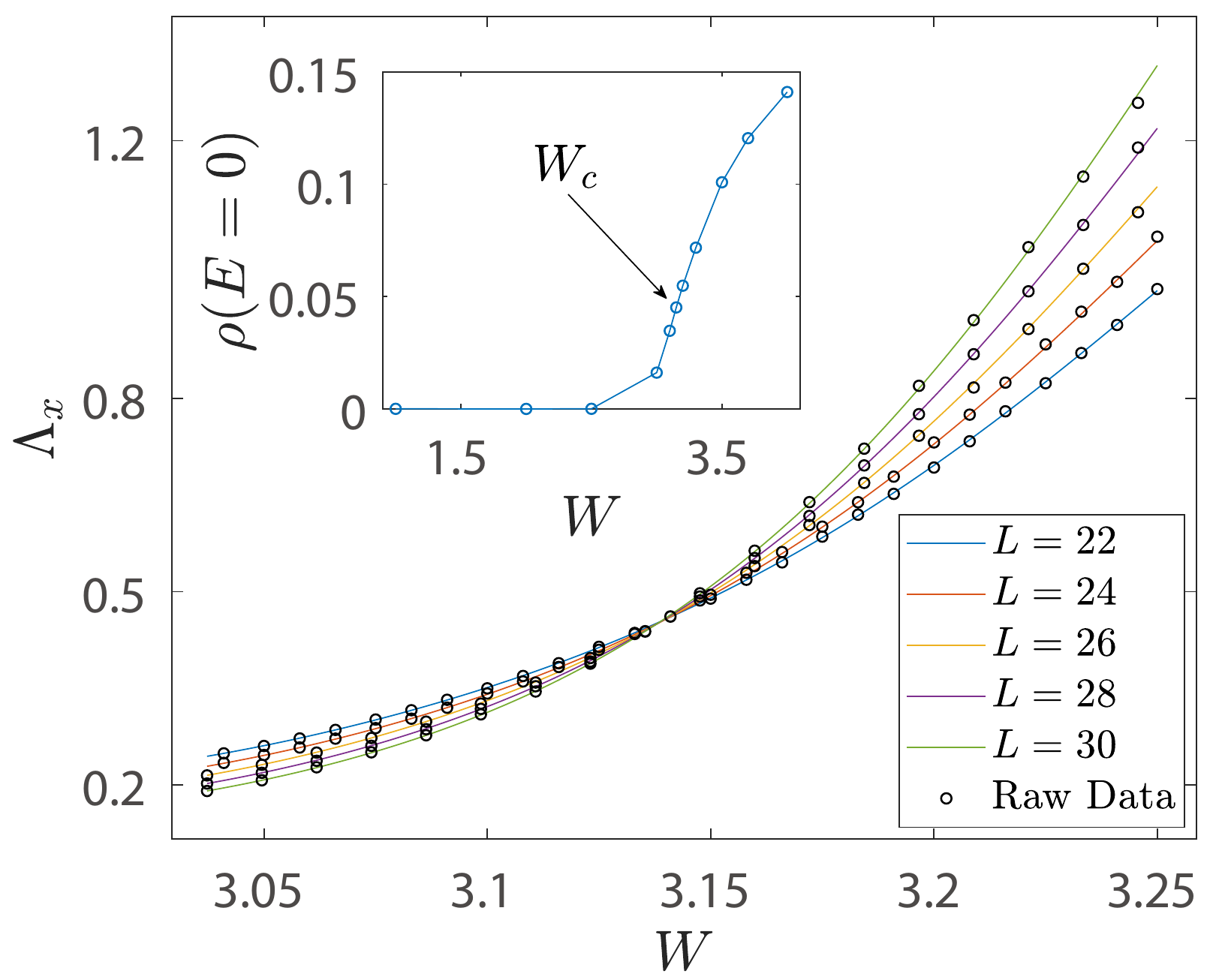}
	\caption{$\Lambda_x\equiv\lambda_x/L$ as a function of the disorder strength around the 
        re-entrant transition between topological band insulator and diffusive metal phases 
        in the 3D class BDI model (see phase transition 1 in Fig.~\ref{fig:1}). The circles are 
        the raw data of $\Lambda_x$, where an error bar is smaller than the circle size. 
        The curves are from the polynomial fitting results with $(m_1,n_1,m_2,n_2)$=(3,3,0,1). 
        Inset: bulk density of states at $E=0$ near the re-entrant transition point.}
	\label{fig:2}
\end{figure}

The same conclusion can be reached by self-consistent Born (SCB) analysis. 
The SCB gives a gap equation for the mean-free time of the zero-energy 
eigenstates of ${\mathbb{H}}$, $\tau$, as;
\begin{align}
1 = K \int_{[-\pi,\pi]^3} \frac{d^3 {\bm k}}{(2\pi)^3} \frac{1}{E^2({\bm k})+\tau^{-2}},  
\end{align}
where $K \equiv W^2/12$ and $E^2({\bm k}) \equiv E^2_{\pm}({\bm k})$ given 
above~\cite{supplemental}. The 3D momentum integral in the right hand side can be 
decomposed into a 1D momentum integral along the closed loop, $k_{\parallel}$, and 
2D momentum integral within the cross-sectional plane, ${\bm k}_{\perp}$. 
Since $E({\bm k})$ has the linear dispersion around the node within the plane, the 2D integral over 
${\bm k}_{\perp}$ has an infrared (IR) logarithmic singularity for any $k_{\parallel}$. Thus, 
the right hand side essentially reduces to $K \int_{1/v_F\tau } dk_{\perp}/k_{\perp}$. 
Due to the IR logarithmic singularity, the gap equation for arbitrary small $K$ 
always leads to a solution of a finite IR `cutoff' $\tau^{-1}$ 
(a finite mean-free time of the zero-energy states), which 
results in a finite zero-energy DOS. In fact, the transfer matrix calculations 
do not indicate the presence of any quantum phase transition 
from NLDSM phase to DM phase at {\it finite} disorder strength (Fig.~\ref{fig:1}).

\textit{Critical exponent in the re-entrant transition}-- 
The critical exponent associated with the Anderson transition between the DM and 
Anderson insulator phases (phase transition 2 in Fig.~\ref{fig:1}) is evaluated 
by the polynomial fitting analysis of the normalized localization length as $\nu=0.80 \pm 0.02$,  
where a localization length along the $x$ direction $\lambda_x$ is calculated~\cite{supplemental}. 
The critical exponent associated with the Anderson transition between the trivial band 
insulator and DM phases (phase transition 3 in Fig.~\ref{fig:1}) is evaluated 
as $\nu=0.82 \pm 0.01$. Since the 95 $\%$ confidence intervals 
of these two exponents (see Table~\ref{table:1}) overlap with each other,
 we conclude that the critical exponent of the Anderson transition in the 
3D class BDI is $\nu=0.80 \pm 0.02$.

Based on this new knowledge, we next evaluate the critical exponent 
associated with the re-entrant insulator-metal transition between 
topological band insulator and DM phases (phase transition 1 in Fig.~\ref{fig:1}). We 
use the same polynomial fitting analyses for the localization length along the $x$ direction 
$\lambda_x$ with $L_x=3\times 10^6$ for 
$L=22$, and $L_x=2\times 10^6$ for $L=24,26,28,30$ (Fig.~\ref{fig:2}). 
The fitting result with GOF greater than 0.1 gives $\nu=0.83 \pm 0.05$ (Table.~\ref{table:1}). 
From the comparison with the other two exponents from the phase transitions 2 and 3, we conclude that 
the re-entrant transition between the topological band insulator and DM phases is of the same universality class 
as the Anderson transition in 3D class BDI. Note also that the KPE calculation shows a finite 
zero-energy DOS on the re-entrant phase transition point (inset of Fig.~\ref{fig:2}). 
The situation is  similar to previous studies on 2D symplectic class, where the quantum spin Hall insulator 
to DM transition shows the same critical exponent as in standard Wigner-Dyson (WD) universality classes
~\cite{Obuse07,Fu12} and to a previous study on 3D unitary class, where the layered Chern insulator 
to DM transition shows the same critical exponent as in the WD universality 
classes~\cite{Luo_quantumMulti}. 

\textit{Summary} --- The critical exponents of Anderson transitions in 3D class CI and 
that in class BDI are clarified numerically.  A disorder-driven re-entrant transition from a topological 
band insulator phase to diffusive metal phase is studied in a model of the class BDI. A comparison 
of the critical exponents suggests that the re-entrant transition belongs to 
the same universality class as the Anderson transition in the class BDI. 
The transfer-matrix calculation as well as self-consistent Born study suggests that an 
infinitesimally small disorder drives the nodal line Dirac semimetal  in the clean limit to 
the diffusive metal.

\textit{Acknowledgment} --- This work (X. L., B. X. and R. S.) was supported by NBRP of China Grants No. 2014CB920901, No. 2015CB921104, and No. 2017A040215.
T. O. was supported
by JSPS KAKENHI Grants No. JP15H03700 and 19H00658.

\bibliography{paper}

\newpage

\section{supplemental materials}
\subsection{3D class CI model}
\begin{figure}[h]
	\centering
	\includegraphics[width=0.95\linewidth]{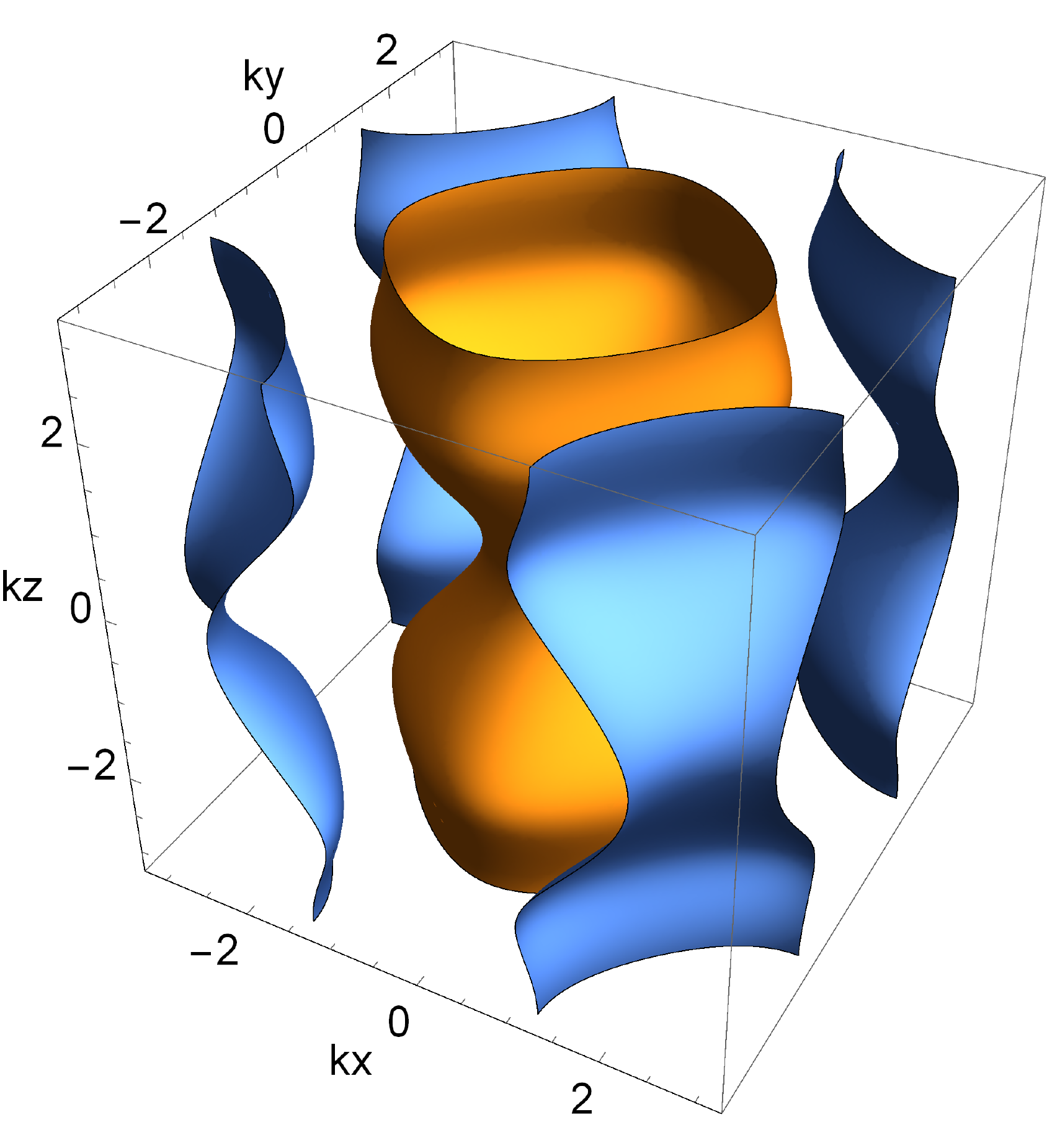}
	\caption{two Fermi surfaces at $E=0$ in the 3D class CI model.}
	\label{sfig:1}
\end{figure}

\begin{figure}
	\centering
	\includegraphics[width=0.95\linewidth]{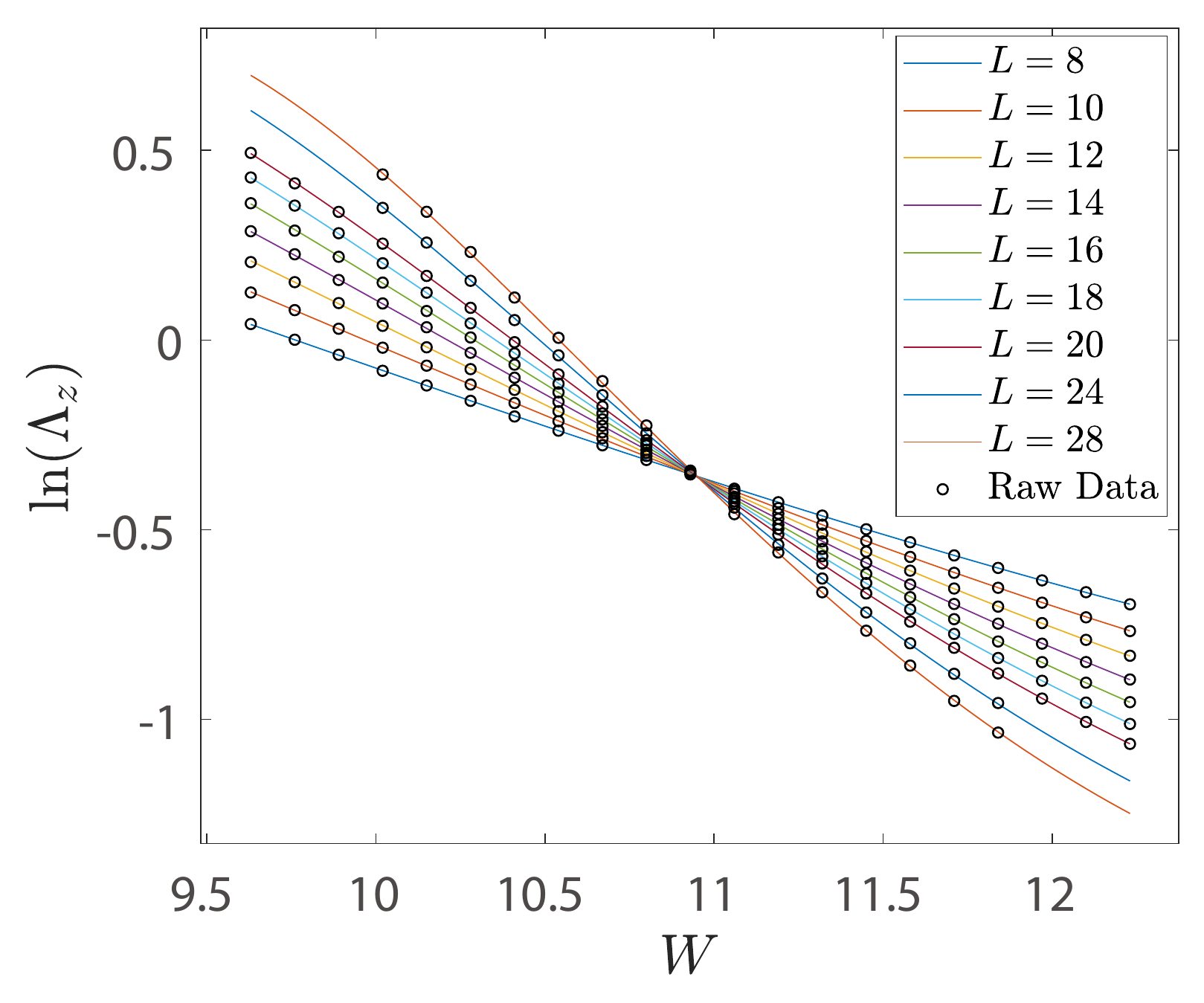}
	\caption{ $\ln(\Lambda_z)\equiv\ln(\lambda_z/L)$ as a function of the disorder strength $W$ for the 
3D class CI model. The circles are the raw data of $\ln(\Lambda_z)$, whose error bar is smaller than the circle 
size. The curves are from the fitting results with the Taylor-expansion 
orders:$(m_1,n_1,m_2,n_2)$=(2,3,0,1).}
	\label{sfig:1a}
\end{figure}

\begin{figure}
	\centering
	\includegraphics[width=1\linewidth]{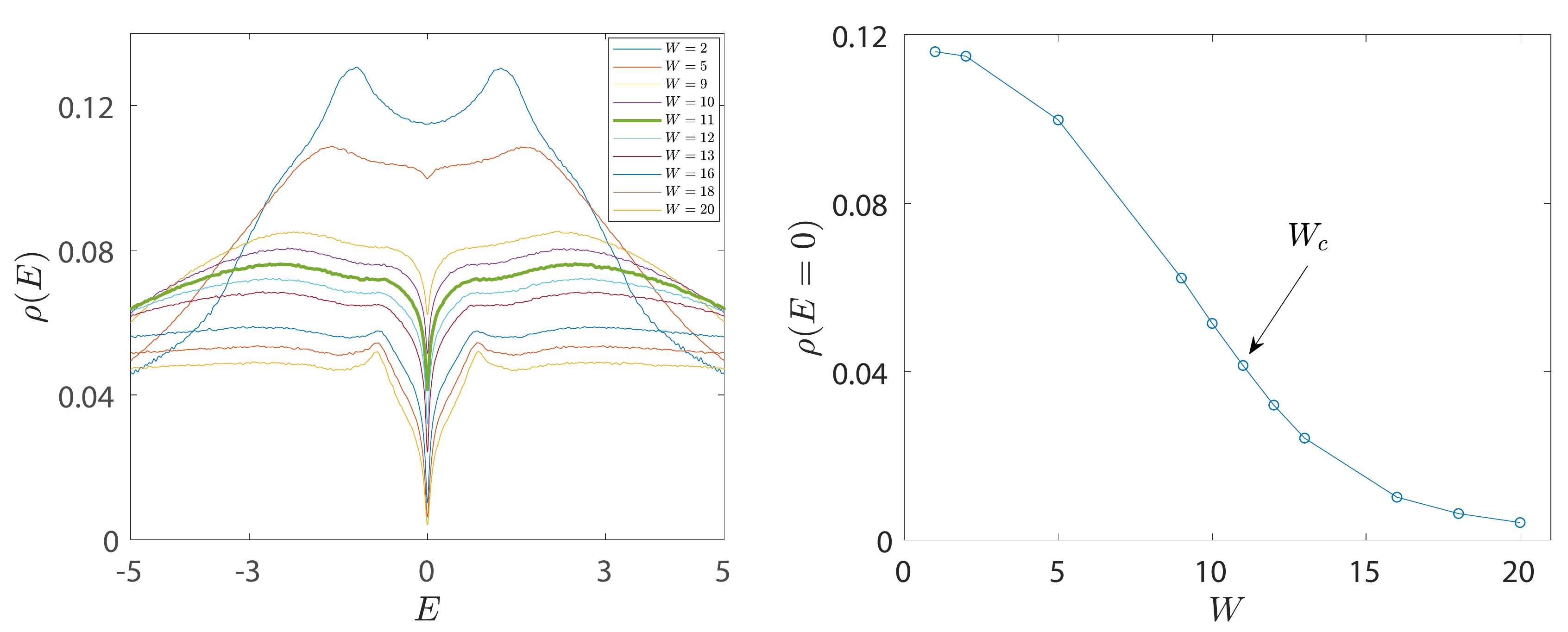}
	\caption{(left) density of states (DOS) as a function of $E$ for several different values of 
         the disorder strength $W$. (right) The zero-energy DOS as a function of the disorder strength.
		The density of states is calculated by the kernel polynomial expansion method~\cite{Weisse06}
		with a cubic system size $L=100$ and large polynomial expansion order ($N$=4000). The DOS is 
         averaged over 20 different disorder realizations.}
	\label{sfig:2}
\end{figure}

The tight-binding model Hamiltonian for the 3D class CI model in the clean limit reduces to 
the following two by two Hamiltonian in the momentum space, 
\begin{align}
{\bm H}({\bm k})= 2t_{\perp} (\cos k_x + \cos k_y) \sigma_0 + t_{\parallel} \sigma_1 
+ (2t^{\prime}_{\parallel} \cos k_z + \Delta) \sigma_3, 
\end{align}
with two separate energy bands, 
\begin{align}
E_{\pm}({\bm k}) = 2t_{\perp} (\cos k_x + \cos k_y) \pm \sqrt{t^{2}_{\parallel} + (2t^{\prime}_{\parallel} \cos k_z + \Delta)^2}.  
\end{align}
In the main text, we set $\Delta=t_{\parallel}=t^{\prime}_{\parallel}=t_{\perp}=1$, where the zero-energy states comprise 
of two disconnected Fermi surfaces (Fig.~\ref{sfig:1}). In the presence of the random potential, the localization length 
of the zero-energy eigenstates ($\lambda_z$) is calculated along $z$ direction as a function of disorder strength $W$, 
where the periodic boundary condition is imposed along $x$ and $y$ directions.  A linear dimension of a cross-section 
of the cubic lattice within the $xy$ plane ($L$) and a linear dimension along the $z$ direction ($L_z$) are set to 
$L_z=2\times 10^6$ for $L=24,28$ and $L_z=4\times 10^6$ for $L=8,10,\cdots,18,20$, respectively. 
On increasing the disorder strength, the zero-energy eigenstates undergo the Anderson transition, where 
the critical disorder strength is identified by a scale-invariant point of the normalized localization length 
$\Lambda_z \equiv \lambda_z/L$ (Fig.~\ref{sfig:1a}). The density of states (DOS) of the system 
with the random pontential is also calculated as a function of the disorder strength $W$ in terms of the 
Kernel polynomial expansion (KPE) method~\cite{Weisse06} for the same set of the tight-binding parameters 
(Fig.~\ref{sfig:2}). For any $W$, the total DOS is an even function in $E$ due to the particle-hole 
symmetry, while a tiny odd component in $E$ stems from finite expansion order in the KPE method. The zero-energy 
DOS decreases on increasing the disorder strength $W$, while it remains finite at the Anderson transition point 
($W_c \simeq 11$; right panel of Fig.~\ref{sfig:2}). 

\subsection{3D class BDI model}
The tight-binding Hamiltonian for 3D class BDI model in the clean limit is given by the following 
2 by 2 Hamiltonian,  
\begin{align}
{\bm H}({\bm k}) =  a_2({\bm k}) \sigma_2 + a_3({\bm k}) \sigma_3 
\end{align}
with 
\begin{align}
a_{2}({\bm k}) &=  2t_{\parallel} \sin k_z  ,  \nonumber \\
a_{3}({\bm k}) &= \Delta + 2t_{\perp} (\cos k_x + \cos k_y) 
+ 2t^{\prime}_{\parallel} \cos k_z.  \nonumber 
\end{align}
\subsubsection{self-consistent Born analyses in favor for single-particle Green function}
A single-particle retarded/advanced Green function is averaged over the random potential $\varepsilon_{\bm i}$;
\begin{align}
&[{\bm G}_{\pm}(E,{\bm k},{\bm k}^{\prime})]_{c,c^{\prime}} = \nonumber \\
& \ \ \frac{1}{N} \sum_{{\bm i},{\bm j}} 
e^{i{\bm k}{\bm i} - i{\bm k}^{\prime}{\bm j}} \bigg\langle 
\bigg[\frac{1}{E-{\mathbb{H}}_{0}-{\mathbb{V}}\pm i\delta}\bigg]_{({\bm i},c|{\bm j},c^{\prime})} \bigg\rangle_{\rm imp}
\end{align}
where $c,c^{\prime}$ denotes the index for the two orbitals, $c,c^{\prime}=a,b$. ${\mathbb{H}}_0$ is 
${\mathbb{H}}$ without the random potential and ${\mathbb{V}}$ is the random potential part,
\begin{align}
\big[{\mathbb{V}}\big]_{({\bm i},c|{\bm j},c^{\prime})} \equiv 
\delta_{{\bm i},{\bm j}} [\sigma_3]_{c,c^{\prime}} \varepsilon_{\bm i}. 
\end{align} 
$\langle \cdots \rangle_{\rm imp}$ stands for a quenched average over the short-ranged random potential and 
is defined below;
\begin{align}
\langle \varepsilon_{\bm i} \varepsilon_{\bm j} \rangle_{\rm imp} &= \delta_{{\bm i},{\bm j}} 
\bigg( \int^{W/2}_{-W/2} \varepsilon^2 d\varepsilon \bigg) \Bigg/ 
\bigg(\int^{W/2}_{-W/2} d\varepsilon \bigg) \nonumber \\
&=  \delta_{{\bm i},{\bm j}} \frac{W^2}{12} 
\equiv \delta_{{\bm i},{\bm j}} K.  
\end{align}
In the thermodynamic limit, the quenched 
average of higher-order powers in the random potential is given by the second-order average, e.g. 
\begin{align}
&\langle \varepsilon_{\bm i} \varepsilon_{\bm j} \varepsilon_{\bm n} \varepsilon_{\bm m} \rangle_{\rm imp} = \nonumber \\
& \ \ (\delta_{{\bm i},{\bm j}} \delta_{{\bm n},{\bm m}} + \delta_{{\bm i},{\bm n}} \delta_{{\bm j},{\bm m}} 
+ \delta_{{\bm i},{\bm m}} \delta_{{\bm j},{\bm n}} ) (K^2 + {\cal O}(N^{-1})).  
\end{align}
The averaged Green function takes a diagonal form in the momentum; $[{\bm G}_{\pm}(E,{\bm k},{\bm k}^{\prime})]_{c,c^{\prime}} 
= \delta_{{\bm k},{\bm k}^{\prime}} [{\bm G}_{\pm}(E,{\bm k})]_{c,c^{\prime}}$, where the two by two 
$[{\bm G}_{\pm}(E,{\bm k})]$ is given by the following Dyson equation within the self-consistent Born approximation,
\begin{align}
{\bm G}_{\pm}(E,{\bm k}) &= {\bm G}_{\pm,0}(E,{\bm k}) \nonumber \\
&\bigg({\bm 1} 
+ \frac{K}{N} \sum_{\bm q} \sigma_3 {\bm G}_{\pm}(E,{\bm q})\sigma_3 \!\ {\bm G}_{\pm}(E,{\bm k})\bigg). \label{scb}
\end{align}
${\bm G}_{\pm,0}(E,{\bm k})$ is the Green function in the clean limit;
\begin{align}
{\bm G}^{-1}_{\pm,0}(E,{\bm k}) &= (E\pm i\delta) - {\bm H}({\bm k})  \nonumber \\
&\equiv a_0 \sigma_0 - a_2({\bm k})\sigma_2 - a_3({\bm k}) \sigma_3 
\end{align}
with $a_0 \equiv E\pm i\delta$. The solution of the Dyson equation is characterized by two ${\bm k}$-independent 
complex-valued constants, $\gamma_0$ and $\gamma_3$;
\begin{align}
{\bm G}^{-1}_{\pm}(E,{\bm k}) = 
\big(a_0 - \gamma_0\big) \sigma_0 - a_2({\bm k})\sigma_2 - \big(a_3({\bm k})+\gamma_3\big) \sigma_3. 
\end{align}
For the zero-energy states ($E=0$), $\gamma_0$ and $\gamma_3$ take pure imaginary and real values respectively;
\begin{align}
{\rm Im}\gamma_0 &= \frac{K}{N} \sum_{\bm q} \frac{{\rm Im}\gamma_0}{a^2_2({\bm q})+ (a_3({\bm q})+{\rm Re}\gamma_3)^2 
+ ({\rm Im}\gamma_0)^2}, \label{gap1} \\
{\rm Re}\gamma_3 &= -\frac{K}{N} \sum_{\bm q} \frac{a_{3}({\bm q}) + {\rm Re}\gamma_3}{a^2_2({\bm q})+ (a_3({\bm q})+{\rm Re}\gamma_3)^2 
+ ({\rm Im}\gamma_0)^2}. \label{gap2}
\end{align}
${\rm Re}\gamma_3$ renormalizes an energy gap in the band insulator phases as well as a shape 
of nodal line in the semimetal phase. 
${\rm Im}\gamma_0$ is an inverse of a mean-free (life) time of the zero-energy states. According to the gap equations, 
${\rm Im}\gamma_0$ can be either zero (`ballistic' zero-energy-states solution) or a finite constant that satisfies Eq.~(\ref{gap2}) 
and 
\begin{align}
1 &= \frac{K}{N} \sum_{\bm q} \frac{1}{a^2_2({\bm q})+ (a_3({\bm q})+{\rm Re}\gamma_3)^2 
+ ({\rm Im}\gamma_0)^2}. \label{gap3} 
\end{align}
Eq.~(\ref{gap3}) corresponds to Eq.~(3) in the main text. The zero-energy density of states is proportional to ${\rm Im}\gamma_0$;
\begin{align}
\rho(E=0) = -\frac{1}{\pi} \frac{1}{N} \sum_{\bm k} {\rm Im}{\rm Tr}\big[{\bm G}_{+}(E=0,{\bm k})\big] 
= \frac{2 {\rm Im}\gamma_0}{\pi K}. \label{dos}
\end{align}
By solving the gap equations numerically, we determine a phase boundary of $\rho(E=0)$; a boundary between 
a phase with $\rho(E=0)=0$ and a phase with $\rho(E=0)\ne 0$ (Fig. 1 in the main text). 
 
\subsubsection{localization length and density of states}
\begin{figure}
	\centering
	\includegraphics[width=1.0\linewidth]{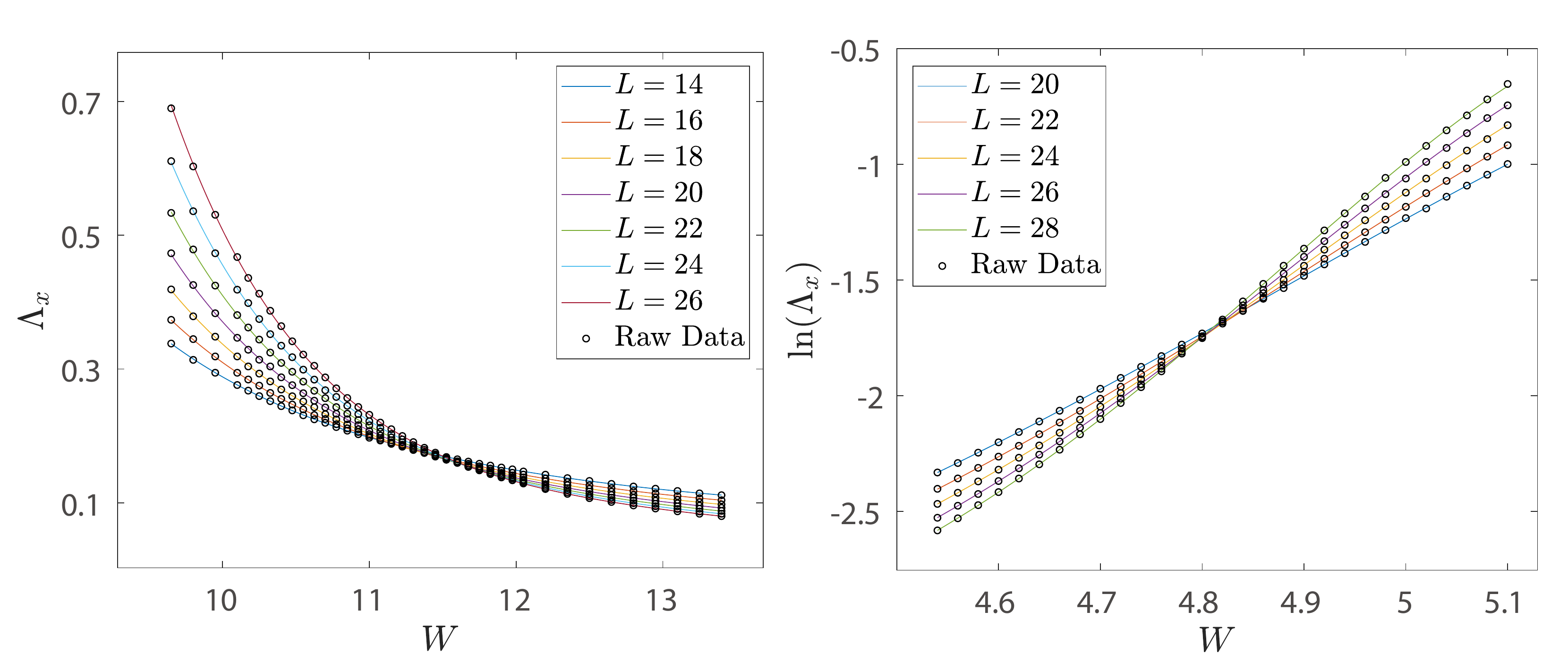}
\caption{Localization length as a function of the disorder strength in the 3D class BDI model. 
  left: $\Lambda_x\equiv\lambda_x/L$ for the phase transition 2 (see in Fig.~1 in the main text).
  right: $\ln(\Lambda_x)\equiv\ln(\lambda_x/L)$  for the phase transition 3 (see in Fig.~1 in the 
  main text). The circles are the raw data, where an error bar is smaller than the circle size. 
  The curves are from the polynomial fitting curves with $(m_1,n_1,m_2,n_2)$=(3,3,0,1) (left) and 
  with $(m_1,n_1,m_2,n_2)$=(2,3,0,1) (right).}
	\label{sfig:3}
\end{figure}

The localization length along $x$ ($\lambda_x$) is calculated as a function of 
disorder strength $W$ at three different parameter points of the 3D class BDI model.
In the calculation, following three quantum phase 
transition points $W_c$ are identified with scale-invariant points of the normalized localization 
length $\Lambda_x \equiv \lambda_x/L$: (i) phase transition 1 between the topological band insulator and DM phases; 
$\Delta=0.5$ and $W_c=3.135$ (Fig.~2 in the main text), (ii) phase transition 2 
between the DM and AI phases; $\Delta=0.5$ and $W_c=11.96$ (left panel of 
Fig.~\ref{sfig:3}), and (iii) phase transition 3 between the trivial band insulator and DM phases; 
$\Delta=4.0$ and $W_c=4.76$ (right panel of Fig.~\ref{sfig:3}). Here $L$ and $L_x$ are a linear 
dimension of the cubic lattice system within the $yz$ plane and along $x$ respectively. For the phase 
transition 2, the localization length $\lambda_x$ is calculated with 
$L_x= 2 \times 10^6$ for $L=26$ and $L_x= 1 \times 10^6$ for 
$L=14, \!\ 16, \!\ 18, \!\ 20, \!\ 22, \!\ 24$. For the phase transition 3, $\lambda_x$ 
is calculated with $L_x= 3 \times 10^6$ for $L=22$ and $L_x= 2\times 10^6$ for 
$L=24, \!\ 26, \!\ 28, \!\ 30$. From the polynomial fitting analyses, the critical exponent of 3D calss BDI 
as well as the MI transition points $W_c$ are precisely determined (Table~I in the main text).  

\begin{figure}
	\centering
	\includegraphics[width=0.95\linewidth]{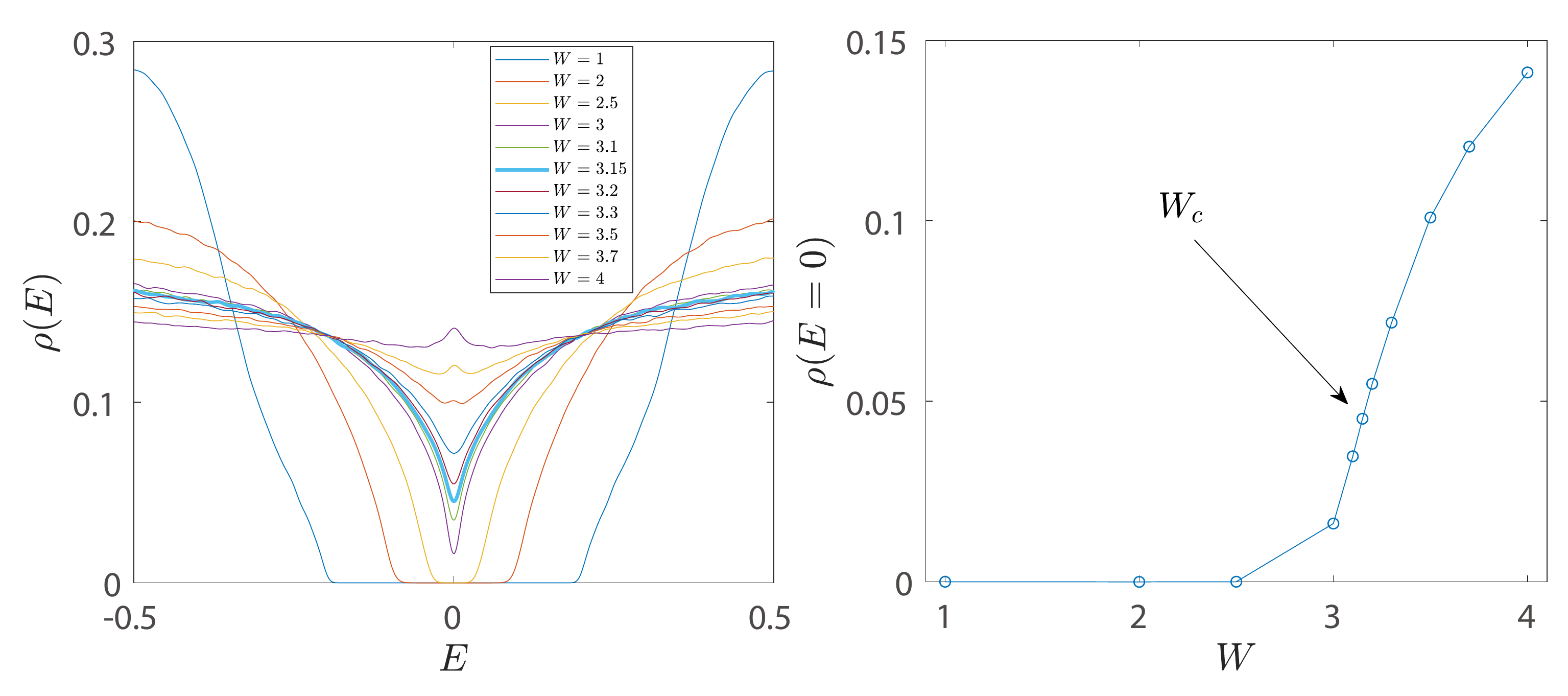}
	\includegraphics[width=0.95\linewidth]{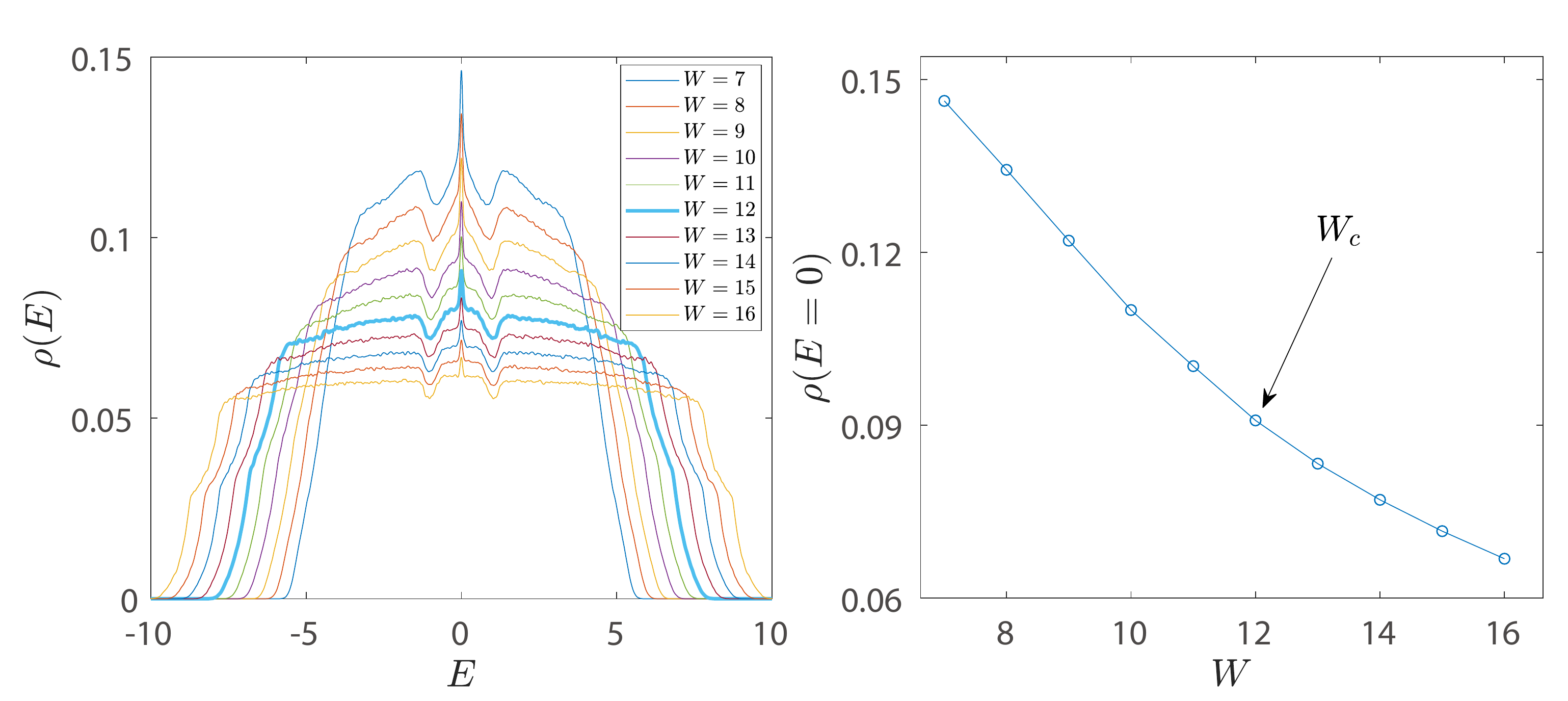}
	\includegraphics[width=0.95\linewidth]{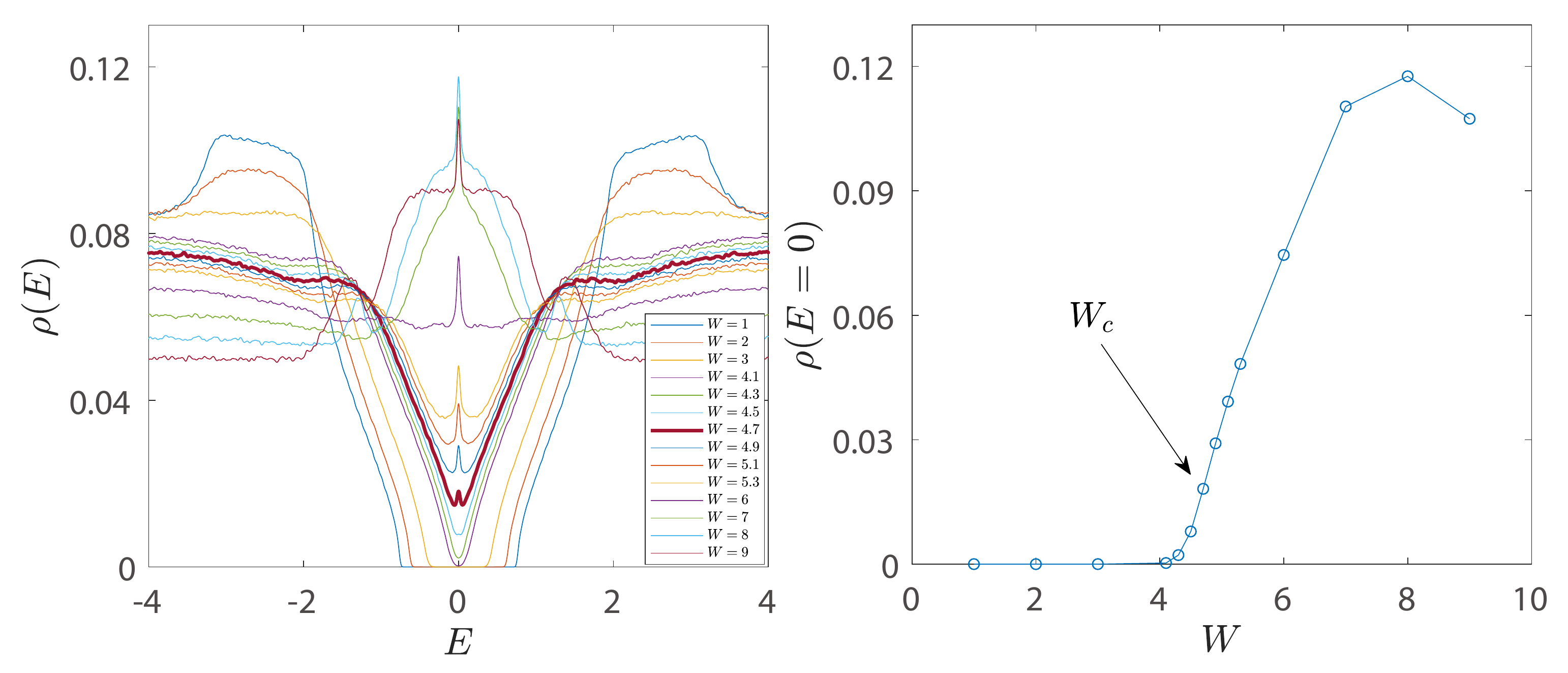}
	\caption{Density of states (DOS) by the kernel polynomial expansion 
                 method~\cite{Weisse06} with the cubic system size $L=100$ around the three quantum 
                 phase transition points. upper: DOS for the phase transition 1 with the polynomial expansion order 
                 $N$=3000; middle:DOS for the phase transition 2 with the polynomial expansion order $N$=2000;
                  lower:DOS for the phase transition 3 with the polynomial expansion order $N$=4000. (Left) 
                 DOS as a function of $E$ and (right) the zero-energy DOS as a function of the disorder strength $W$.  
	             We take an average over four different disorder realizations with fixed $W$. The bold line in the left  
                 figures indicate a disorder strength that is closest to $W_c$.}
	\label{sfig:4}
\end{figure}

The DOS is also calculated for the same sets of parameters in terms of the KPE method 
(Fig.~\ref{sfig:4}). The zero-energy DOS is always finite at the MI transition points of 
the three phase transitions. Especially for the phase transitions 1 and 3, the zero-energy 
DOS becomes finite at a certain critical disorder strength below $W_c$. 
The critical disorder strengths for the phase transitions 1 and 3 are 
consistent with the boundary determined by the self-consistent Born analyses. 

\subsubsection{one dimensional limit}
When $t_{\perp}=0$, the 3D class BDI model reduces to a one-dimensional (1D) model;
\begin{align}
{\bm H}({\bm k}) &= 2t_{\parallel} \sin k_z \sigma_2 + (\Delta + 2 t^{\prime}_{\parallel} \cos k_z) \sigma_3 \nonumber \\
&= \sqrt{(2t_{\parallel} \sin k_z )^2 + (\Delta + 2t^{\prime}_{\parallel} \cos k_z)^2} \nonumber \\
& \hspace{2cm} \times \big(n_2(k_z) \sigma_2 + n_3(k_z) \sigma_3\big). \label{1d}
\end{align}
The topological integer for the 1D BDI topological insulator is defined as a winding number of the two-component 
unit vector $(n_2(k_z),n_3(k_z))$ as a function of $k_z \in [-\pi,\pi]$;
\begin{align}
{\mathbb{Z}} \equiv \int^{\pi}_{-\pi} \frac{dk_z}{2\pi} \Big( n_3 
\partial_{k_z} n_2 - n_2 \partial_{k_z} n_3 \Big). \label{wn}
\end{align}
When $|\Delta| < 2|t^{\prime}_{\parallel}|$ with $t_{\parallel} \ne 0$, the integer is $\pm 1$, while the integer is zero for 
$|\Delta| > 2|t^{\prime}_{\parallel}|$ with $t_{\parallel} \ne 0$.  The topological integers of  
the topological phase in Fig.~1 in the main text are $+1$ for any $k_x$ and $k_y$. 

\begin{figure}
	\centering
	\includegraphics[width=0.95\linewidth]{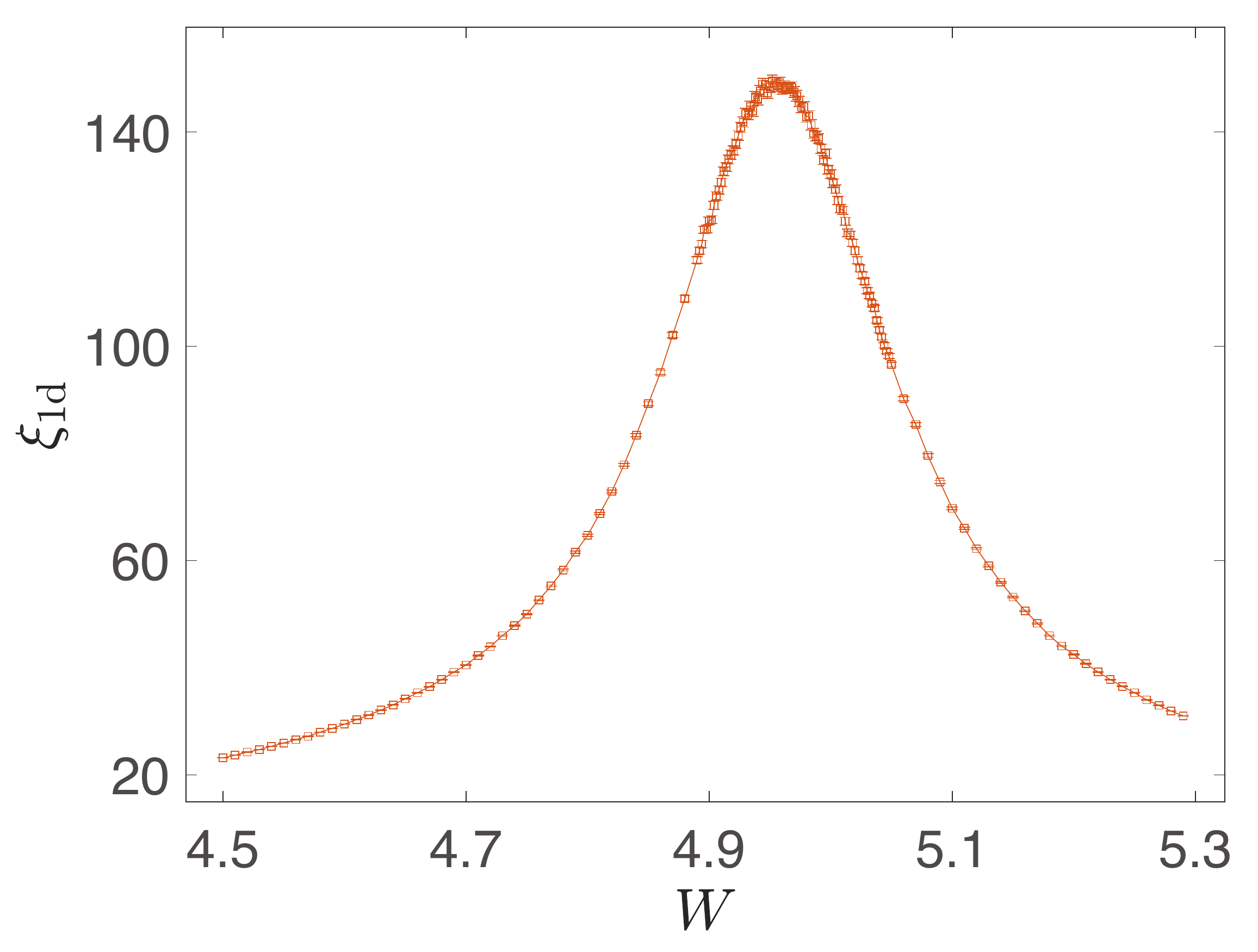}
	\caption{1D localization length $\xi_\mathrm{1d}$ as a function of the disorder strength $W$ for 
     $\Delta=t_{\perp}=0$, $t_{\parallel}=-1$, and $t^{\prime}_{\parallel}=-1/4$.
	1D BDI topological insulator phase is separated from the Anderson insulator at a point around $W\simeq 5.0$, 
    where the localization length shows a very strong peak.}
	\label{sfig1d}
\end{figure}

When the random potential is weakly introduced with the BDI symmetry, the topological integer 
remains unchanged, unless the zero-energy bulk states become delocalized. 
On the one hand, the bulk eigenstates in the strongly disorder regime must be in a conventional 
localized phase with the zero topological integer. This suggests that between the 1D BDI topological 
insulator phase in the weakly disordered regime and 1D conventional localized phase in the strongly disordered regime, 
there must be a insulator to insulator transition at a certain disorder strength. 
To test this numerically, we set $t_\perp=0$ and calculate a 1D localization length $\xi_\mathrm{1d}$ for
$\Delta=0, t_\parallel'=-1$ and $t_\parallel=-1/4$ (Fig.~\ref{sfig1d}). The localization length shows a very 
strong peak around $W\simeq 5.0$, indicating a certain transition from 1D topological insulator phase 
to Anderson insulator phase. The point named as `1d limit' in Fig.~1 in the main text is determined by 
the value of $W$ at which $\xi_{\rm 1d}$  shows the sharp peak. Numerically, however, the peak value remains 
finite even for very large $L_z \simeq 10^8$. We leave it for future study a detailed behaviour of $\xi_{\rm 1d}$ 
in this one-dimensional limit.  

\end{document}